\documentclass[10pt, aps, pra, superscriptaddress,  twocolumn]{revtex4-2}
\pdfoutput=1

\usepackage[english]{babel}
\usepackage[autostyle=true]{csquotes}
\usepackage{natbib}
\usepackage{amsmath}
\usepackage{amssymb}
\usepackage{physics}
\usepackage{braket}
\usepackage{dcolumn}
\usepackage{graphicx}
\usepackage{multirow}
\usepackage{makecell} 
\usepackage[none]{hyphenat}
\usepackage[normalem]{ulem}
\usepackage{parskip}
\usepackage{CJK}
\usepackage{soul}
\usepackage[dvipsnames]{xcolor}
\newcommand{\beginsupplement}{%
        \setcounter{table}{0}
        \renewcommand{\tablename}{Supplementary Table}%
        \setcounter{figure}{0}
        \renewcommand{\figurename}{Supplementary Fig.}%
        \setcounter{equation}{0}
        \renewcommand{\theequation}{S\arabic{equation}}%
        \setcounter{section}{0}
        \renewcommand{\thesection}{\arabic{section}} 
}
\usepackage[colorlinks=true, linkcolor=black, citecolor=black, urlcolor=black]{hyperref}
\hypersetup{CJKbookmarks=true} 
\usepackage{bookmark} 
\usepackage[final]{microtype}

\begin{document}
\title{Tailoring spatial correlations with quantum interference}

\author{Carlo Schiano}
\author{Bereneice Sephton}
\affiliation{Dipartimento di Fisica, Universit\`a di Napoli Federico II, Complesso Universitario di Monte S. Angelo, Via Cintia, 80126 Napoli, Italy}
\author{Elnaz Darsheshdar}
\affiliation{Dipartimento di Fisica, Universit\`a di Napoli Federico II, Complesso Universitario di Monte S. Angelo, Via Cintia, 80126 Napoli, Italy}
\author{Lorenzo Marrucci}
\affiliation{Dipartimento di Fisica, Universit\`a di Napoli Federico II, Complesso Universitario di Monte S. Angelo, Via Cintia, 80126 Napoli, Italy}
\affiliation{CNR-ISASI, Institute of Applied Science and Intelligent Systems, Via Campi Flegrei 34, 80078 Pozzuoli (NA), Italy}
\author{Corrado de Lisio}
\author{Vincenzo D'Ambrosio}
\email[]{vincenzo.dambrosio@unina.it}
\affiliation{Dipartimento di Fisica, Universit\`a di Napoli Federico II, Complesso Universitario di Monte S. Angelo, Via Cintia, 80126 Napoli, Italy}

\begin{abstract}
Photon correlations represent a central resource in many quantum optics experiments, with applications ranging from quantum information protocols to sensing. Engineering such correlations is often challenging, especially in multi-particle scenarios. In this work we describe an effective method for shaping spatial correlations between photons by patterning their distinguishability in a quantum interference setup. We show how to write and edit these bi-photon correlations between the two output channels of a beam-splitter, hiding this encoded information from conventional intensity measurements. Our scheme offers an easy extension to multiparticle scenarios and facilitates the transmission of high-dimensional quantum information, with potential applications to quantum communication and imaging protocols. 
\end{abstract}

\maketitle

\section{Introduction}

Light is a versatile physical system that plays a central role in both fundamental and applied research. 
A direct measurement of light's properties, such as intensity or wavelength, enables powerful techniques like microscopy and spectroscopy, with applications ranging from  medicine to astrophysics. Beyond direct measurements, further advancements are possible by analysing correlations in light, overcoming some of the typical limitations in standard photonic schemes  \cite{lubin2022photon, Defi24}. 
Today, correlations are a ubiquitous workhorse in quantum photonics experiments, which facilitates applications such as fundamental tests of quantum mechanics \cite{bigbelltest} and the development of quantum protocols ranging from quantum imaging \cite{gilaberte2023experimental,genovese2016real} to enhanced detection \cite{england2019quantum} and communication \cite{jackson2002optical}.  
Developing ways to tailor such correlations is therefore a crucial step for photonic technologies within this framework.

Most commonly, correlations are  sourced as bi-photons from spontaneous parametric down-conversion (SPDC) where predominant customisation focuses on tailoring the pump laser and non-linear crystal (NLC) characteristics.
 For instance, NLCs are engineered through nonlinearity profile modulation \cite{Branczyk11, Dosseva2016, Yesh23} or phase-matching \cite{Valencia2007} and the pump modes are either given complex structures or amplitude-modulated by introducing objects into their path \cite{Mair2021, Kovlakov2017, Kovlakov2018, Boucher2021, Unternahrer:18, Fran21, Defienne2024}.
Alternatively, each photon's structure can be augmented post-creation, typically by placing geometric phase elements, such as q-plates (QPs) \cite{DAmb16, Gao_2024} or spatial light modulators (SLMs) \cite{Cameron2024} in its path. 
Beyond two-photons, however, progress has been slow in this framework, owing to experimental complexity, limiting advances to just three photons \cite{3xSPDC}.

Outside of SPDC, photonic correlations can be created, for instance, through Hong-Ou-Mandel (HOM) interference. Here two indistinguishable photons, overlapped on a symmetric beam splitter (BS), either bunch into the same output port or anti-bunch into separate ports, depending on their joint state symmetry \cite{bouchard2020two}.
This phenomenon has been used to produce global correlations (averaged along the transverse plane) between photons by controlling their indistinguishability across degrees of freedom such as polarization \cite{kwiat1992observation} or spatial modes \cite{walborn2003multimode,nagali2009optimal,di2010measurement,karimi2014exploring}. 
More recently, transitioning from global to local, structured light in a quantum eraser setting has enabled the heralding of spatially patterned correlations between one output port of the BS and a bucket detector, by locally manipulating the indistinguishability across each photons' spatial profile \cite{Schi2024}. Moreover, quantum interference between structured light modes has been exploited to generate the complete Bell basis in the polarisation pattern of two photons \cite{Gao25}.

Here, we extend this approach to show that it is possible to locally tailor spatially resolved correlations between both output ports of the HOM BS. We do this by tuning interference conditions point-by-point, in the transverse plane, with spatially varying polarization modes. In addition, polarization projections after the BS can be incorporated to further edit these structures. 
Our results enable the transmission of complex, high-dimensional information, encoded in photonic correlations, with potential applications in quantum technologies such as quantum communication and imaging. Furthermore, they offer additional insight into the spatial structure of quantum interference, revealing significant local variations beyond the global averages typically considered \cite{bornman2019ghost,zhang2017simultaneous,zhang2016engineering,Casalengua_2024_VortexCorrelations}. While demonstrated here for two photons, our approach can be readily extended, proffering a way forward for engineering non-local spatial correlations in multi-photon scenarios. 

\begin{figure*}
    \centering
    \includegraphics[width=0.95\linewidth]{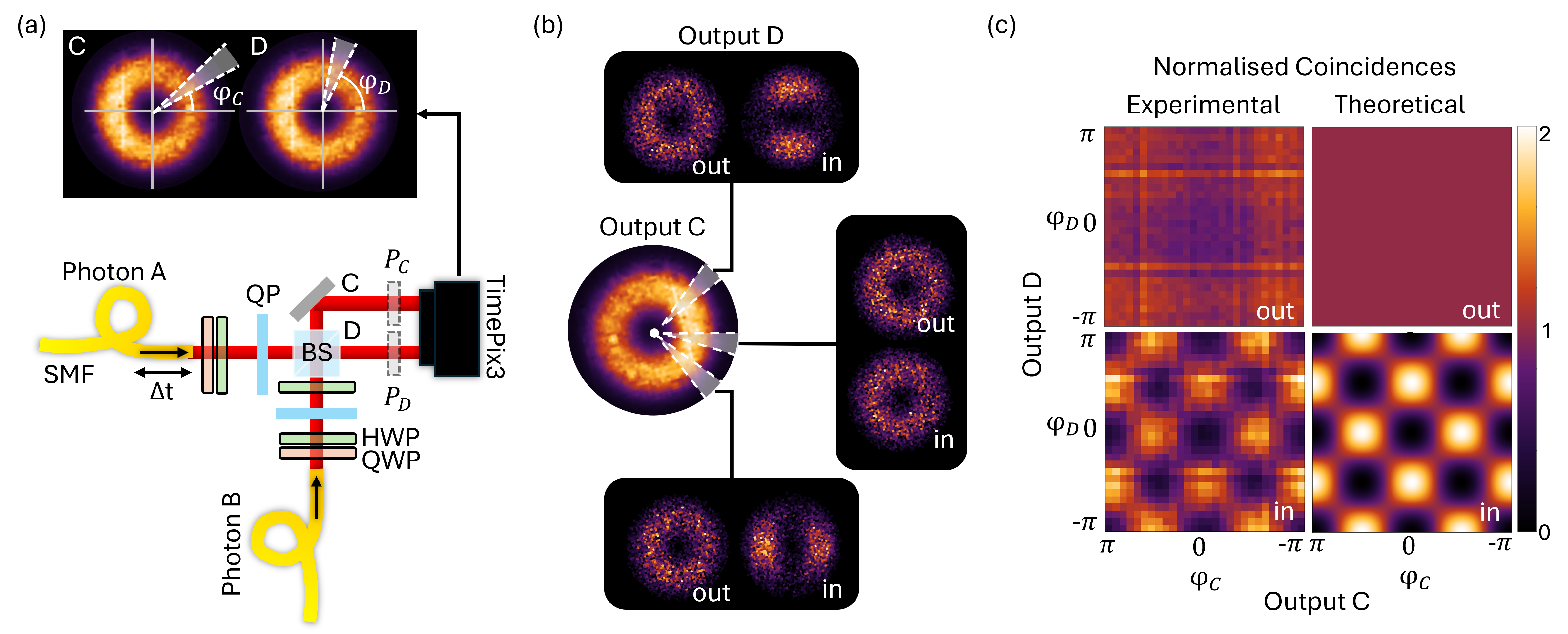}
    \caption{\textbf{Tailoring structured correlations.} (a) Experimental setup used to write correlations by mixing two spatially structured photons with a 50:50 beam splitter. Top inset depicts an example of the 
    selected regions of photon detection events for coincidence measurement in the angular coordinate ($\varphi_C, \varphi_D$) of the transverse plane, for each output port. (b) Angular dependence of the experimental correlations between ports C and D when the two photons are temporally indistinguishable (in) and distinguishable (out). In this case input modes are the azimuthally-varying VV modes described in Eqs.~\ref{Eq_VVrad} and \ref{Eq_VVpi}. The angular regions in port C heralding the full spatial distributions in port D are highlighted and superimposed on the full distribution of singles (center). 
    (c) Full mapping of coincidences in the angular coordinates between both output ports for the data in (b), showing azimuthal dependency of the correlations of temporally matched photons (in) in contrast to a featureless distribution observed when the photons are temporally distinguishable (out).
    The coincidence distributions are normalised with respect to the temporally distinguishable (out) configuration. SMF: single mode fiber; QWP: quarter-waveplate; HWP: half-waveplate, QP: $q$-plate; BS: 50:50 beam-splitter; $P_C,P_D$: polariser.}
    \label{Fig1}
\end{figure*}

\section{Concept}
\noindent

We focus on the case where the two input photons are in a separable state:
\begin{equation}
\ket{\psi}_{in} 
= 
a^\dagger_{\sigma_A, \tau_A}
b^\dagger_{\sigma_B,\tau_B}
\ket{0}
\end{equation}
where
$a^{\dagger}_{\sigma_A,\tau_A}$ and $b^{\dagger}_{\sigma_B,\tau_B}$
are creation operators for the structured modes $\sigma_A$ and $\sigma_B$, propagating along paths A and B,  associated with time-bins $\tau_A$ and $\tau_B$ respectively, and acting on the vacuum state.
By sending the two photons into different input ports of a 50:50 BS and post-selecting the cases where they exit from separate output ports, $C$ and $D$, the resulting quantum state is:\\ 
\begin{align}
|\Psi\rangle_{out} = \frac{1}{2} \left[c^\dagger_{\sigma_A, \tau_A} d^\dagger_{\sigma_B, \tau_B} -  c^\dagger_{\sigma_B, \tau_B} d^\dagger_{\sigma_A, \tau_A} \right] \ket{0}~.
\label{eq: two-photons state - out}
\end{align}
Here, $c^\dagger$ and $d^\dagger$ are creation operators corresponding to the output modes at the beam splitter ports $C$ and $D$, respectively. 

Under the paraxial approximation, each input mode can be expressed as:
\begin{equation}\label{eq: field mode}
\mathbf{E}(\mathbf{r}, t) = \mathbf{e}_\sigma(\mathbf{r}) f (\mathbf{r}, t - \tau) e^{-i\omega t}~,
\end{equation}
where $\omega$ is the carrier frequency, $f(\mathbf{r},t)$ denotes the mode spatial profile, and $\mathbf{e}_{\sigma}(\mathbf{r})$ is a position-dependent transverse unit vector that specifies both the local polarisation state and the phase profile of the mode (see SI for a detailed discussion).
For simplicity, we assume that $f(\mathbf{r},t)$ is identical for the two input photons and that the vector $\mathbf{e}_\sigma$ is constant along the propagation axis. This can therefore be expressed in terms of its position-dependent components, $H_\sigma(\mathbf{r}_\perp)$ and $V_\sigma(\mathbf{r}_\perp)$,  as:
\begin{align}\label{eq: e_vector}
\mathbf{e}_\sigma(\mathbf{r}_\perp)
=
\mathbf{e}_H H_\sigma(\mathbf{r}_\perp) 
+
\mathbf{e}_V
V_\sigma(\mathbf{r}_\perp)~,
\end{align}
where $\mathbf{r}_\perp$ is the transverse position vector, and $\mathbf{e}_H$ and $\mathbf{e}_V$ are unit vectors denoting horizontal and vertical linear polarisations, respectively.

We are interested in correlations between detection events in positions $\mathbf{r}_{C\perp}$ and $\mathbf{r}_{D\perp}$ defined in two transverse planes (detection planes) along output modes C and D, respectively.  
The coincidence probability distribution $C_{\tau_A, \tau_B}(\mathbf{r}_{C \perp}, \mathbf{r}_{D \perp})$ for the two output photons, can be obtained by 
calculating the fourth-order correlation function \cite{Photons_Atoms} and integrating over the detection time windows for detectors placed in each output port (see SI for details). 
In the following, we will always consider transverse positions, therefore we will omit the subscript $\perp$ to simplify the notation.
We discriminate between temporally indistinguishable photons ($ \tau_A = \tau_B$, referred to as \textit{in}), where quantum interference can take place: 
\begin{align}\label{eq: coincidence probability - in}
    C_{(In)}(\mathbf{r}_C; \mathbf{r}_D) = &\\ \nonumber 
    A(\mathbf{r}_C, \mathbf{r}_D)\sum_{\{\alpha, \beta\}}&
    \left\vert
    \left[ \mathbf{u}_{\alpha} \cdot \mathbf{e}_{\sigma_A}(\mathbf{r}_{C}) \right]
    \left[ \mathbf{u}_{\beta} \cdot \mathbf{e}_{\sigma_B}(\mathbf{r}_{D}) \right]
    \right. \\ \nonumber
    &\left.
    -
    \left[ \mathbf{u}_{\alpha} \cdot \mathbf{e}_{\sigma_B}(\mathbf{r}_{C}) \right]
    \left[ \mathbf{u}_{\beta} \cdot \mathbf{e}_{\sigma_A}(\mathbf{r}_{D}) \right]
    \right\vert^2~,
\end{align}
and temporally distinguishable photons ($\tau_A \neq \tau_B$, referred to as \textit{out}), where no quantum interference takes place:
\begin{align}\label{eq: coincidence probability - out}
    C_{(Out)}(\mathbf{r}_C; \mathbf{r}_D)& = \\ \nonumber
    A(\mathbf{r}_C, \mathbf{r}_D)\sum_{\{\alpha, \beta\}}&
    \{ \left\vert
    \left[ \mathbf{u}_{\alpha} \cdot \mathbf{e}_{\sigma_A}(\mathbf{r}_{C}) \right]
    \left[ \mathbf{u}_{\beta} \cdot \mathbf{e}_{\sigma_B}(\mathbf{r}_{D}) \right]
    \right\vert^2  \\ \nonumber
    &
    +
    \left\vert
    \left[ \mathbf{u}_{\alpha} \cdot \mathbf{e}_{\sigma_B}(\mathbf{r}_{C}) \right]
    \left[ \mathbf{u}_{\beta} \cdot \mathbf{e}_{\sigma_A}(\mathbf{r}_{D}) \right]
    \right\vert^2 \}~,
\end{align}

where $A(\mathbf{r}_C, \mathbf{r}_D)$ is a factor that depends on the input mode spatial profile (see SI for more details).
The sums in Eqs.~(\ref{eq: coincidence probability - in})~and~(\ref{eq: coincidence probability - out}) are taken over all possible polarisation configurations \(\{\mathbf{u}_{\alpha}, \mathbf{u}_{\beta}\}\) of an orthonormal polarisation basis.
To quantify the correlations introduced  when moving from temporal matching (in) to temporal mismatch (out) between photons, we use the visibility function \cite{weihs1996two}, defined as: 
\begin{equation}\label{eq: visibility}
    \mathcal{V}(\mathbf{r}_C; \mathbf{r}_D)=
    \frac{
    C_{out}(\mathbf{r}_{C}; \mathbf{r}_{D})-C_{in}(\mathbf{r}_{C}; \mathbf{r}_{D})
    }{
    C_{out}(\mathbf{r}_{C}; \mathbf{r}_{D})
    }~,
\end{equation}
which takes values ranging from -1 (anti-bunching) to +1 (bunching).
Crucially, the spatial dependence of the visibility function is directly determined by the choice of $H_\sigma(\mathbf{r_\perp})$ and $V_\sigma(\mathbf{r}_\perp)$ in Eq.~\eqref{eq: e_vector} for the two input modes.

These correlations can be further edited by performing polarisation projections in each output port, therefore providing an extra degree of control.
This can be calculated from Eq.~\eqref{eq: visibility} by eliminating the sum over the independent polarisation directions (See SI) and choosing the polarisation unit vectors, $\mathbf{u}_{\alpha}$ and $\mathbf{u}_{\beta}$, to be aligned with the projective measurement directions.

\section{Results}

\begin{figure}
    \centering
    \includegraphics[width=1\linewidth]{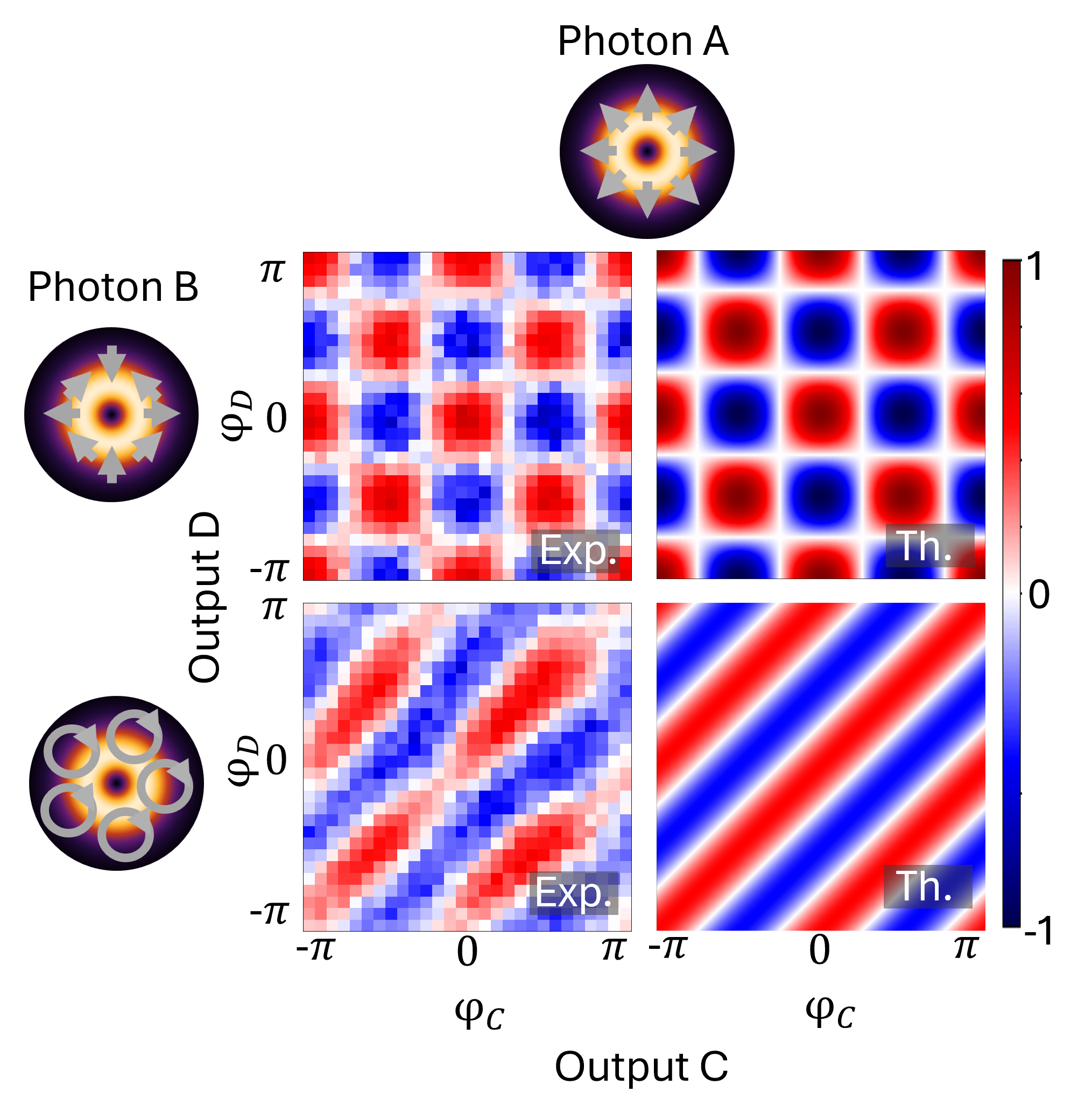}
    \caption{\textbf{Visibility distributions.} Experimental (Exp.) and theoretical (Th.) visibility distributions showing the correlations written by structuring photon A with a radial VV mode (top row inset) and photon B (left column insets) with a $\pi-$VV mode (first row) and circularly polarised, $\ell = 1$ scalar mode (second row).
    }
    \label{Fig2}
\end{figure}

\noindent We experimentally demonstrate our scheme with the setup illustrated in Fig.~\ref{Fig1}(a) (a detailed description is provided in the SI). Two orthogonally polarised photons (photons A and B) with a wavelength of $\approx 810$ $nm$ are each initialized into TEM00 spatial mode through single mode fibres (SMF). Each photon is then prepared in the desired structured mode by using birefringent waveplates (HWP, QWP) and a q-plate (QP)  before being directed onto different input ports of a 50:50 beam splitter (BS). An additional half-waveplate in arm B allows one to further modify the polarisation structure of photon B. The temporal mismatch $\Delta t$ between the two photons is controlled by a delay line in the path of photon A.
Both BS output paths, C and D, are directed onto separate regions of a single-photon event camera, serving as independent regions for the measurement of spatial correlations between the two photons. This is illustrated in the top inset of Fig.~\ref{Fig1}(a), where each output region is assigned coordinates centred on the transverse distribution.
Although, in principle, the spatial correlations span a four-dimensional domain defined by the transverse coordinates of each pixel, we restrict our analysis to angular correlations by employing modes that depend only on the azimuthal coordinate $\varphi$ in the transverse plane.
Such symmetry allows us to integrate over the radial coordinate, simplifying the visualisation of the results.

Accordingly, we choose to tailor our photons with the azimuthally-varying vector vortex (VV) mode structure,
\begin{equation}
    \mathbf{e}_{rad}(\varphi) = 
    \cos{\varphi}~ \mathbf{e}_{H} + \sin{\varphi}~ \mathbf{e}_{V}~,
    \label{Eq_VVrad}
\end{equation}
in arm A and $\pi$-VV mode 
\begin{equation}
    \mathbf{e}_{\pi}(\varphi) = 
    \cos{\varphi}~ \mathbf{e}_{H} - \sin{\varphi}~ \mathbf{e}_{V}~,
    \label{Eq_VVpi}
\end{equation}
in arm B, prepared by impinging the QPs with photons having uniform horizontal polarisation \cite{DAmb16}. 
In doing so, photon correlations are azimuthally patterned between the two BS outputs as a result of the varying local quantum interference conditions. 
Fig.~\ref{Fig1}~(b)~and~(c) show this patterning when the two photons are temporally indistinguishable (in) in contrast to the unstructured distribution when they are distinguishable (out). In particular, Fig.~\ref{Fig1}~(b) depicts how the spatial distribution of the photons detected in port D changes when heralded in coincidence with selected angular positions in port C. We highlight three prominent cases as an example with the angular positions superimposed on the singles distribution detected in port C. Here, quantum interference gives rise to vertical (top) and horizontal (bottom) lobes when photons are temporally tuned, while certain positions yield no structure (middle).  Fig.~\ref{Fig1}~(c) shows the full mapping of both ports' angular dependencies for the correlations written by the VV modes interference. The coincidences observed when the two photons are temporally tuned (bottom row) reveal an azimuthal dependence in the local quantum interference, contrasting with a homogeneous distribution when temporally distinguishable (top row), in good agreement with the expected distributions. 

A clearer picture of these tailored correlations can be seen by considering the spatially varying visibility  (Eq.~\ref{eq: visibility}), shown in the upper row of Fig.~\ref{Fig2}.  
The chequered visibility distribution, $\mathcal{V}(\varphi_{C}; \varphi_{D}) = \cos{2\varphi_{C}}\cos{2\varphi_{D}}$ alternates between photon bunching (red, corresponding to a HOM dip) and anti-bunching (blue, corresponding to a HOM peak) with a periodicity of $\pi$ (see SI for detailed calculations). Interestingly, as a result of their orthogonality ($\mathbf{e}_{rad} \cdot \mathbf{e}_{\pi} = 0 $),  spatial correlations arising from these modes can only be observed if both coordinates ($\varphi_{C}$ and $\varphi_{D}$) are experimentally accessible. Otherwise, bucket detection in one or both of the two arms, $\mathcal{V}(\varphi_{C})=\int{\mathcal{V}(\varphi_{C};\varphi)d\varphi=0}$ and $\mathcal{V}(\varphi_{D})=\int{\mathcal{V}(\varphi;\varphi_{D})d\varphi=0}$, reveals no structure.

As discussed, we can easily manipulate the correlation pattern by selecting different input states. For example, photon B can be set in the mode:\\ 
\begin{equation}
    \mathbf{e}^{(l)}_{\circlearrowright}(\varphi) = \frac{1}{\sqrt{2}}
    (
    \mathbf{e}_{H} + e^{i\frac{\pi}{2}}\mathbf{e}_{V}
    )
    e^{il\varphi}~,
    \label{Eq_OAM}
\end{equation}
by switching its polarization to circular, before impinging the QP.
For such a mode, each photon carries OAM of \(l\) (in units of $\hbar$), where, in our experiment, $l=1$.
Keeping photon A unchanged, the correlations now resemble diagonal stripes in visibility where $\mathcal{V}(\varphi_{C};\varphi_{D}) = \frac{1}{2}\cos{(\varphi_{C}-\varphi_{D})}$, which is confirmed experimentally (Fig.~\ref{Fig2}, lower row).  A reduction in the visibility bound to $\pm\frac{1}{2}$ results from the inherent distinguishability limit placed by the mutually unbiased relation between the circular and linear polarisation distributions of the two overlapping photon structures. In general, the periodicity ($\frac{2\pi}{(\ell_A - \ell_B)}$) can be controlled by the net difference between the OAM imparted to photon A ($\ell_A$) and B ($\ell_B$) where the polarisation is indistinguishable.

\begin{figure*}
    \centering
    \includegraphics[width=0.85\linewidth]{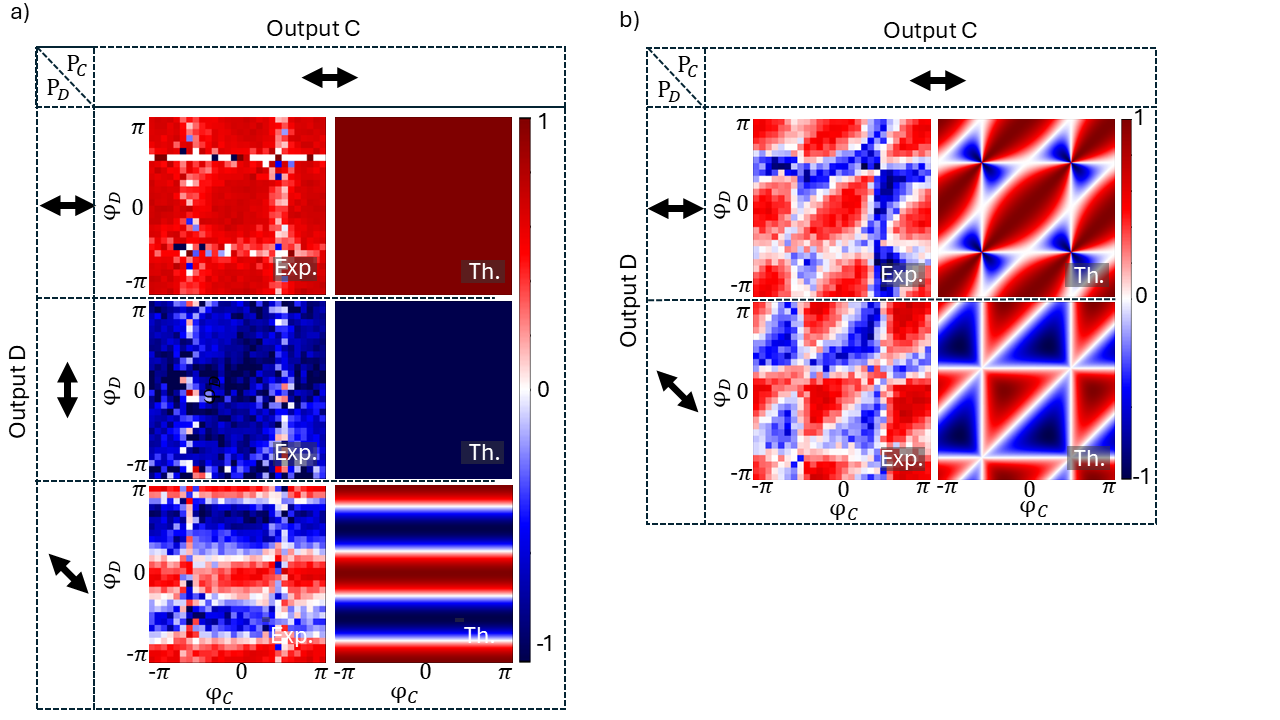}
    \caption{\textbf{Editing correlations with polarisation erasure.}  Experimental (Exp.) and theoretical (Th.) visibility distributions showing edited correlations by performing polarisation projections when input photon A is in a radial VV mode and photon B in a (a) $\pi-$VV mode and a (b) circularly polarised $\ell = 1$ scalar mode. Polariser orientations are indicated by the arrows in the top and left sections of the tables.  
    }
    \label{Fig4}
\end{figure*}

Interestingly, writing correlations by structuring the polarisation degree of freedom, also enables an easy \enquote{editing} by erasing the locally-defined polarisation distinguishability \cite{kwiat1992observation,Schi2024}. This is achieved by means of including polarisers ($P_C$, $P_D$) in each output arm (C, D) as indicated in Fig.~\ref{Fig1} (a). A sample set of measurements with horizontal (H), vertical (V) and anti-diagonal (A) polarisation projections are presented in Fig.~\ref{Fig4}. In particular, Fig.~\ref{Fig4} (a), reports the modified correlations for input radial and $\pi-$VV modes using projections P$_C$ $\in$ \{H,V,A\} and P$_D$ $\in$ \{H\}, showing good agreement between the experimental and expected cases. By selecting polarization projections, it is possible to extract a range of outcomes from uniform bunching (Fig.~\ref{Fig4} (a) top row) and anti-bunching (middle row) to dependence on only one of the output angular coordinates ($\phi_D$, bottom row). Alternatively, for the input modes described by Eqs.~\ref{Eq_VVrad}~and~\ref{Eq_OAM}, more complex spatial correlations can be composed by using projections $P_C$ $\in$ \{H,A\} and $P_D$ $\in$ \{H\}. This is shown in Fig.~\ref{Fig4} (b), with visibilities,
\begin{equation}
\mathcal{V}(\varphi_{C};\varphi_{D}) = \frac{2\cos{\varphi_{C}}\cos{\varphi_{D}}\cos(\varphi_{C}-\varphi_{D})}{\cos^{2}{\varphi_{C}} + \cos^{2}{\varphi_{D}}
}
\end{equation}
featuring a bow-tie spatial correlation pattern (top row) and 
\begin{equation}
\mathcal{V}(\varphi_{C};\varphi_{D}) = 
\frac{\cos{2\varphi_{D}} - \sin{2\varphi_{C}}+\cos{2(\varphi_{C}-\varphi_{D})}}{2+\cos{2\varphi_{C}} -\sin{2\varphi_{D}}}
\end{equation}
with a triangular spatial correlation pattern (bottom row). 
Here, fair agreement between the experimental and analytical correlation patterns can be observed, with slight asymmetries in the experimental distributions creating some distortions in the patterning.

\section{Discussion and conclusion}
\noindent
Although we have demonstrated correlation editing for a bi-photon arrangement, our scheme can be easily extended to incorporate additional photons, providing access to more complex structured correlations \cite{Gard:13}. In contrast, such scalability with SPDC processes is more demanding. There, the maximum number of correlated photons generated by a single SPDC process is limited to three \cite{3xSPDC}, and further advancements in this field require technological progress in  source engineering. Alternatively, the generation of multiphoton correlations through the entanglement of multiple SPDC sources \cite{12-Photon-Entanglement} is a widely used approach. However, this method faces significant challenges, including the need for a large number of SPDC sources and relatively low coincidence rates. Structuring multiphoton correlations through quantum interference, as demonstrated in our experiment, is a powerful alternative with successful  demonstrations in the path degree of freedom \cite{20xBoson_Sampling}. A key advantage of this approach is that it relies solely on photon indistinguishability, regardless of the type of source used to generate them. Moreover,  the control of photon structures in the transverse plane paves the way for higher-dimensional correlations and greater flexibility in their editing. We note that a challenge of this approach may lie in the synchronization of the interfering beams, where additional sensitivity to propagation-related Gouy phases from different modes should be kept in mind \cite{jaouni2025tutorial}.
Furthermore, we note that while we have demonstrated our scheme using the azimuthal degree of freedom, this can be easily applied to more complex distributions, such as arbitrary complex images. 
Additionally, by incorporating dynamic elements to structure the photons, such as spatial light modulators instead of static $q$-plates, correlations could be  edited on demand for a versatile source of correlations.

In conclusion, we have introduced a scheme, based on quantum interference, to write arbitrary spatially-dependent correlations between two photons. 
These correlations are tailored such that neither ordinary intensity measurements, nor lack of spatial resolution in the case of orthogonal modes, reveals the written structure. This allows high-dimensional information to be encoded in photon correlations and hidden for secure communication purposes, using a simple setup and readily-available elements. Furthermore, by exploiting the polarisation degree of freedom, our technique offers additional tailoring capabilities as projections onto different polarisations edit the correlations and gives another tool in the encoding/decoding process. As our scheme enables the embedded transmission of complex and editable information, it can be useful for developing quantum imaging protocols. Our results also offer insight into localised features in the quantum interference of structured photons impinging a BS, showing that post-selection onto local regions differ substantially from the average or global value, weather detected globally in both outputs \cite{bornman2019ghost,zhang2017simultaneous,zhang2016engineering} or hybridised with spatial detection in one arm \cite{Schi2024}. This can have implications, for instance, in understanding additional localised features when using HOM interference in schemes such as editing spatial entanglement in OAM \cite{zhang2016engineering} or entanglement swapping \cite{zhang2017simultaneous}, among others.

%%%%%%%%REFERENCES%%%%%%%%%%%%

\nocite{Photons_Atoms}
\nocite{Intro_QED}

\section*{ }
\noindent

\noindent \textbf{Acknowledgements:} We would like to thank Prof. Fabrice P. Laussy for helpful and stimulating discussions. The authors acknowledge financial support from the Italian Ministry of Research (MUR) through the PRIN 2022 project “A quantum neuromorphic recognition machine of quantum states” (QNoRM), the PNRR MUR project PE0000023-NQSTI and the PNRR MUR project CN 00000013-ICSC.

\noindent \textbf{Data and materials availability:} Data underlying the results presented in this paper may be obtained from the authors upon reasonable request.

\newpage \clearpage

\setcounter{page}{1}

\onecolumngrid
\begin{center}
    \textbf{\Large Supplementary Information: Tailoring spatial correlations with quantum interference}

    \vspace{0.5 cm}

    Carlo Schiano,$^1$ Bereneice Sephton,$^{1}$ Elnaz Darsheshdar,$^{1}$ Lorenzo Marrucci,$^{1,2}$ Corrado de Lisio,$^1$ and Vincenzo D'Ambrosio$^1$

\vspace{0.5 cm}
    
$^1$Dipartimento di Fisica, Universit\`a di Napoli Federico II, Complesso Universitario di Monte S. Angelo, Via Cintia, 80126 Napoli, Italy

$^2$ CNR-ISASI, Institute of Applied Science and Intelligent Systems, Via Campi Flegrei 34, 80078 Pozzuoli (NA), Italy

\end{center}

\vspace{0.5 cm}

\onecolumngrid

\beginsupplement{

\section{Structured correlations formulation} 

In this section, we provide a detailed theoretical framework for the results presented in the main text. We consider two photons in structured modes as inputs to A and B ports of a 50:50 beam splitter (BS) and measure two-fold coincidences arising from the quantum interference between the two modes. 
The resulting spatial structures vary depending on whether a polarization projection is applied, the specific type of projection used, and the choice of input beam structure. The transverse component of the electric field operator is given by \cite{Photons_Atoms}:
\begin{equation}\label{eq: transverse field superposition}
\hat{\mathbf{E}}_{\perp}(\mathbf{r},t) = \hat{\mathbf{E}}^{+}_{\perp}(\mathbf{r},t) + \hat{\mathbf{E}}^{-}_{\perp}(\mathbf{r},t)~,
\end{equation}
where, the positive and negative frequency part of the transverse field are expressed as:
\begin{equation}\label{eq: transverse field components}
\begin{aligned}
&\hat{\mathbf{E}}^{+}_{\perp}(\mathbf{r},t) = \sum_n \mathbf{E}_n(\mathbf{r},t) \hat{a}_n~, \\ 
&\hat{\mathbf{E}}^{-}_{\perp}(\mathbf{r},t) = \sum_n \mathbf{E}^*_n(\mathbf{r},t) \hat{a}^\dagger_n~.
\end{aligned}
\end{equation}
Here, $\{\mathbf{E}_n(\mathbf{r},t)\}$ is a complete set of orthonormal modes, where $\hat{a}_n$, $\hat{a}_n^{\dagger}$ are the annihilation and creation operators of photons in mode $n$ and satisfy the commutation relation $[\hat{a}_n, \hat{a}_{n'}^{\dagger}] = \delta_{n,n'}$. 
For convenience, we assume that these modes occupy a finite spatial volume $\Omega$ and satisfy the appropriate boundary conditions, thus discretizing the mode set.

\subsection{No Polarisation Projection} 
Let's first consider the case where no polarisation projection is performed.
For any given quantum state $\ket{\psi}$, the number of double detections per unit of time between photons hitting two separate broad-band detectors is proportional to the fourth-order correlation function \cite{Intro_QED} and is given by: 
\begin{equation}
    w^{(II)}(\mathbf{r}_{1}, t_{1}; \mathbf{r}_{2}, t_{2}) = 
    \sum_{\{\alpha, \beta\}}
    \bra{\Psi}
    \left[\mathbf{u}_{\alpha}\cdot\hat{\mathbf{E}}_{\perp}^{-}(\mathbf{r}_{1}, t_{1})\right]  
    \left[\mathbf{u}_{\beta}\cdot\hat{\mathbf{E}}_{\perp}^{-}(\mathbf{r}_{2}, t_{2})\right] 
    \left[\mathbf{u}_{\beta}\cdot\hat{\mathbf{E}}_{\perp}^{+}(\mathbf{r}_{2}, t_{2})\right]  
    \left[\mathbf{u}_{\alpha}\cdot\hat{\mathbf{E}}_{\perp}^{+}(\mathbf{r}_{1}, t_{1})\right]
    \ket{\Psi}~,
\label{eq: fourth-order correlation function}
\end{equation}
where the sum is extended over a set of independent polarisations, such as $\{$Horizontal $(H)$, Vertical $(V)\}$ or $\{$ Left~Circular$(L)$, Right~Circular$(R)\}$.
The unit-vectors $\mathbf{u}_{\alpha}$ and $\mathbf{u}_{\beta}$ represent the direction of the polarisation and can generally be complex.

To proceed, we select a mode basis consisting of vector beams characterized by eigenstates of orbital angular momentum (OAM) with eigenvalues $l$ ($L_z = l\hbar$), and a radial index $p \geq 0$ representing different orthogonal radial modes for each $|l|$. We focus on quasi-monochromatic pulsed modes centered on frequency $\omega$, associated with two discrete input paths $P \in \{A, B\}$, and distinguished by temporal bins $\tau$ and $\tau'$. Under the paraxial approximation, with $z$ as the propagation axis, these modes take the form:
\begin{equation}\label{eq: final field mode}
\mathbf{E}_n(\mathbf{r}, t) = \mathbf{e}_\sigma f_{|l|,p} (r_P, z_P, t - \tau) e^{il\varphi_P} e^{-i\omega t}~,
\end{equation}
where $(r_P , z_P , \varphi_P )$ are the cylindrical coordinates for path $P$, $f_{|l|,p}$ defines the radial amplitude profile, and $\mathbf{e}_\sigma$ is the polarization unit vector perpendicular to the propagation axis $z$. Hence, a mode is uniquely identified by the tuple $n = (P, l, p, \omega, \tau, \sigma)$. The orthogonality of modes with different indices is assumed.
Rather than describing the vector beams as superpositions of modes like in Eq.(\ref{eq: final field mode}), it is also possible (and in our case simpler) to introduce in Eq.(\ref{eq: final field mode}) a position-dependent transverse unit vector that specifies
both the local polarisation state and the phase profile
of the mode,
$\mathbf{e}_{\sigma}(r,\varphi)$, as below:
\begin{align}\label{eq:6}
\mathbf{e}_\sigma(r,\varphi) = \mathbf{e}_H H_\sigma(r,\varphi) + \mathbf{e}_V V_\sigma(r,\varphi)~,
\end{align}
where $\mathbf{e}_H$ and $\mathbf{e}_V$ are unit vectors denoting horizontal and vertical polarisation, respectively.
Here, we are assuming $\mathbf{e}_\sigma$ to be constant along the propagation axis and $H_\sigma(r,\varphi)$ and $V_\sigma(r,\varphi)$ to be arbitrary position dependent functions that satisfy the normalization condition.

We now consider a general quantum state describing two photons entering the BS via input ports A and B. All properties of the photons, except the spatially varying polarisations denoted by the subscripts $n$ and $n'$ and the temporal bins $\tau_A$ and $\tau_B$, are assumed to be identical
to ensure a position-dependent indistinguishability when $\tau_A = \tau_B$. The input state is:

\begin{equation}\label{eq: input state}
\ket{\psi}_{in} = \sum_{n, n'} 
\psi_{n, n'} 
a^\dagger_{n, \tau_A}
b^\dagger_{n', \tau_B}
\ket{0}~,   
\end{equation}

The indices $n$ and $n'$ span a complete set of orthonormal, spatially-dependent polarisation mode.
Here, $a^{\dag}_{n, \tau_A}$ and $b^{\dag}_{n', \tau_B}$ are the position-dependent creation operators for structured modes $n$ and $n'$, propagating along A and B, and associated with time-bins $\tau_A$ and $\tau_B$, respectively. 
The function $\psi_{n, n'}$ denotes the joint two-photon wave packet.
Here, we focus on the simplest case, where the two photons entering the BS from separate ports are initially in a product state $\psi_{n, n'} = \delta_{n, \sigma_A}\,\delta_{n', \sigma_B}$.
Thus, we have $\ket{\psi}_{in} = a^\dagger_{\sigma_{A}, \tau_A} b^\dagger_{\sigma_{B}, \tau_B} \ket{0}$. 

By considering a 50:50 BS acting on the input two-photon state as follows,
\begin{equation}
\begin{aligned}
    &a^\dagger_{\sigma,\tau} = (c^\dagger_{\sigma,\tau} + d^\dagger_{\sigma,\tau})/\sqrt{2}~, \\ 
    &b^\dagger_{\sigma,\tau} = (c^\dagger_{\sigma,\tau} - d^\dagger_{\sigma,\tau})/\sqrt{2}~,
\end{aligned}
\end{equation}
the quantum state $\ket{\Psi}$ in Eq.(\ref{eq: fourth-order correlation function}) can be obtained as:
\begin{equation}\label{eq: BSoutput}
\ket{\Psi}= \frac{1}{2} \left[ c^\dagger_{\sigma_A, \tau_A} d^\dagger_{\sigma_B, \tau_B} - c^\dagger_{\sigma_B, \tau_B} d^\dagger_{\sigma_A, \tau_A} \right] \ket{0}~.   
\end{equation}
Here, $c^\dagger$ and $d^\dagger$ are creation operators for the respective modes at the BS output ports C and D, associated with time bins $\tau_A$ and $\tau_B$. To obtain Eq. (\ref{eq: BSoutput}), we post-selected the states in which the two photons are separated in ports C and D. 
Substituting (\ref{eq: BSoutput}) and (\ref{eq: final field mode}) into (\ref{eq: fourth-order correlation function}) and applying standard commutation relations leads to the fourth-order correlation function:
\begin{align}\label{eq: double detections rate - two photons state - OAM modes}
w^{(II)}_{\tau_A, \tau_B} (\mathbf{r}_C, t_C; \mathbf{r}_D, t_D) = \frac{1}{4}\sum_{\{\alpha, \beta\}} & \Big| [\mathbf{u}_\alpha \cdot \mathbf{e}_{\sigma_A}(r_C,\varphi_C) f(r_C, z_C, t_C-\tau_A)] [\mathbf{u}_\beta \cdot \mathbf{e}_{\sigma_B}(r_D,\varphi_D) f(r_D, z_D, t_D-\tau_B)] - \\ \nonumber
& [\mathbf{u}_\alpha \cdot \mathbf{e}_{\sigma_B}(r_C,\varphi_C) f(r_C, z_C, t_C-\tau_B)] [\mathbf{u}_\beta \cdot \mathbf{e}_{\sigma_A}(r_D,\varphi_D) f(r_D, z_D, t_D-\tau_A)] \Big|^2~,
\end{align}
where $\mathbf{r}_C$ and $\mathbf{r}_D$ are position vectors along the detection plane of output ports $C$ and $D$.

In order to get the coincidence probability between the two BS outputs,  the correlation function must be integrated over the relevant detection window in both $t_1$ and $t_2$:
\begin{equation}
    C_{\tau_A, \tau_B}(\mathbf{r}_{C}; \mathbf{r}_{D}) = \int w^{(II)}_{\tau_A, \tau_B}(\mathbf{r}_{C}, t_{C}; \mathbf{r}_{D}, t_{D}) dt_{C} dt_{D}~. 
\end{equation} 
We now distinguish between two different temporal configurations of the incoming photons: the case of temporally indistinguishable photons ($ \tau_A = \tau_B$) and the case of temporally distinguishable photons ($\tau_A \neq \tau_B$). 
In the former case, we are reproducing the "in the dip/peak" condition of a typical HOM experiment and obtain:
\begin{align}\label{eq: C_in}
    C_{(In)}(\mathbf{r}_{C}; \mathbf{r}_{D}) = &
    \frac{F(r_{C}, z_{C}) F(r_{D}, z_{D})}{4}  \\ \nonumber
    & \sum_{\{\alpha, \beta\}} \left| [\mathbf{u}_\alpha \cdot \mathbf{e}_{\sigma_A}(r_C,\varphi_C)] 
    [\mathbf{u}_\beta \cdot \mathbf{e}_{\sigma_B}(r_D,\varphi_D)] - 
    [\mathbf{u}_\alpha \cdot \mathbf{e}_{\sigma_B}(r_C,\varphi_C)] 
    [\mathbf{u}_\beta \cdot \mathbf{e}_{\sigma_A}(r_D,\varphi_D)]\right|^2~.
\end{align}
In the latter case, assuming that both time bins fall in the integration window and do not overlap, we are reproducing the "out the dip/peak" condition of a typical HOM experiment and obtain:
\begin{align}\label{eq: C_out}
    C_{(Out)}(\mathbf{r}_{C}; \mathbf{r}_{D}) = &
    \frac{F(r_{C}, z_{C}) F(r_{D}, z_{D})}{4} \\ \nonumber
    &\sum_{\{\alpha, \beta\}}  \left\{ \left| [\mathbf{u}_\alpha \cdot \mathbf{e}_{\sigma_A}(r_C,\varphi_C)] 
    [\mathbf{u}_\beta \cdot \mathbf{e}_{\sigma_B}(r_D,\varphi_D)] \right|^2 + \left| [\mathbf{u}_\alpha \cdot \mathbf{e}_{\sigma_B}(r_C,\varphi_C)] 
    [\mathbf{u}_\beta \cdot \mathbf{e}_{\sigma_A}(r_D,\varphi_D)] \right|^2 \right\}~.
\end{align}
Here, we introduced the ``fluence" function,
\begin{equation}
    F(r, z) = \int\limits_{\substack{\text{integration} \\ \text{window}}} f^{2}(r, z, t - \tau)dt~,
\end{equation}
which can be factorised in Eq.~(\ref{eq: C_in}) and (\ref{eq: C_out}) under the assumption that the two input modes share the same transverse spatial profile.
In the \{H, V\} polarisation basis we have:
\begin{align}\label{eq:19}
C_{(In)}(\mathbf{r}_C; \mathbf{r}_D) = \frac{A(\mathbf{r}_C, \mathbf{r}_D)}{4} & \big\{ \left| H_{\sigma_A}(r_C,\varphi_C) H_{\sigma_B}(r_D,\varphi_D) - H_{\sigma_B}(r_C,\varphi_C) H_{\sigma_A}(r_D,\varphi_D) \right|^2 \\ \nonumber
 & + \left| 
 H_{\sigma_A}(r_C,\varphi_C) V_{\sigma_B}(r_D,\varphi_D) - H_{\sigma_B}(r_C,\varphi_C) V_{\sigma_A}(r_D,\varphi_D) \right|^2 \\ \nonumber
 & + \left| 
 V_{\sigma_A}(r_C,\varphi_C) H_{\sigma_B}(r_D,\varphi_D) - V_{\sigma_B}(r_C,\varphi_C) H_{\sigma_A}(r_D,\varphi_D) \right|^2 \\ \nonumber
 & + \left| 
 V_{\sigma_A}(r_C,\varphi_C) V_{\sigma_B}(r_D,\varphi_D) - V_{\sigma_B}(r_C,\varphi_C) V_{\sigma_A}(r_D,\varphi_D) \right|^2 \big\}~,
\end{align}
and,
\begin{align}\label{eq:20}
C_{(Out)}(\mathbf{r}_C; \mathbf{r}_D) = \frac{A(\mathbf{r}_C, \mathbf{r}_D)}{4} &  \big\{  \left| 
H_{\sigma_A}(r_C,\varphi_C) 
H_{\sigma_B}(r_D,\varphi_D) \right|^2 + \left| 
H_{\sigma_B}(r_C,\varphi_C) 
H_{\sigma_A}(r_D,\varphi_D) \right|^2  \\ \nonumber
& + \left| 
H_{\sigma_A}(r_C,\varphi_C) 
V_{\sigma_B}(r_D,\varphi_D) \right|^2 + \left| 
H_{\sigma_B}(r_C,\varphi_C) 
V_{\sigma_A}(r_D,\varphi_D) \right|^2  \\ \nonumber
& + \left| 
V_{\sigma_A}(r_C,\varphi_C) 
H_{\sigma_B}(r_D,\varphi_D) \right|^2 + \left| 
V_{\sigma_B}(r_C,\varphi_C) 
H_{\sigma_A}(r_D,\varphi_D) \right|^2  \\ \nonumber
& + \left| 
V_{\sigma_A}(r_C,\varphi_C) 
V_{\sigma_B}(r_D,\varphi_D) \right|^2 + \left| 
V_{\sigma_B}(r_C,\varphi_C) 
V_{\sigma_A}(r_D,\varphi_D) \right|^2 \big\}~,
\end{align}
where we set $A(\mathbf{r}_C, \mathbf{r}_D) = F(r_{C}, z_{C}) F(r_{D}, z_{D})$ in order to simplify the notation. To observe the variation in the correlations between the photons when interfered (in) and not (out), we introduce the visibility function \cite{weihs1996two}, defined as
\begin{equation}
    \mathcal{V}(\mathbf{r}_{C}; \mathbf{r}_{D}) = 
    \frac{
    C_{(Out)}(\mathbf{r}_{C}; \mathbf{r}_{D}) - C_{(In)}(\mathbf{r}_{C}; \mathbf{r}_{D})
    }
    {
    C_{(Out)}(\mathbf{r}_{C}; \mathbf{r}_{D})  
    }~.
\end{equation}

We evaluate the visibility function in three cases of interest in the following sections.

\paragraph{Case: {$vv1=rad$ and $vv2=\pi$}.}
\begin{flushleft}
We consider the case where input vector modes are the radial and $\pi$ VV ones. These are respectively described by the following azimuthally varying polarization vectors:
\begin{equation}
\begin{aligned}
    &\mathbf{e}_{rad}(\varphi) = 
    \mathbf{e}_{H}\cos{\varphi} + \mathbf{e}_{V}\sin{\varphi}~, \\ 
    &\mathbf{e}_{\pi}(\varphi) = 
    \mathbf{e}_{H}\cos{\varphi} - \mathbf{e}_{V}\sin{\varphi}~.
\end{aligned}
\end{equation}
Here $e_{i}$, $i \in \{H,V\}$ is the horizontal ($H$) or vertical ($V$) polarisation unit-vector.
The coincidence probability and visibility functions obtained for these modes are: 
\begin{equation}
\begin{aligned}
&C_{(Out)}(\varphi_C,\varphi_D) =  \frac{A(\mathbf{r}_C, \mathbf{r}_D)}{2}~,\\ 
&C_{(In)}(\varphi_C,\varphi_D) = \frac{A(\mathbf{r}_C, \mathbf{r}_D)}{2} (1-\cos{2\varphi_C}\cos{2\varphi_D})~, \\ 
&\mathcal{V}(\varphi_C,\varphi_D) = \cos{2\varphi_C}\cos{2\varphi_D}~.
\end{aligned}
\end{equation}
 In this case the quantum interference is occurring between two orthogonal VV modes ($\mathbf{e}_{vv1} \cdot \mathbf{e}_{vv2} = 0 $).
This implies that the correlations described by this formula can be experimentally observed only if both DoFs, $\varphi_{C}$ and $\varphi_{D}$, are accessible.
When only one of these DoFs is experimentally accessible, the corresponding theoretical visibility function $\mathcal{V} = \mathcal{V}(\varphi)$ is obtained by integrating over the inaccessible DoF. In this case, the visibility function is uniformly null.
\end{flushleft}

\paragraph{Case: $vv1=rad$ and $vv2=(oam1)_\circlearrowright$.}
\begin{flushleft}
We consider the case where the input modes are the radial VV beam and a circular polarised beam carrying Orbital Angular Momentum (OAM) with $l=1$, described by the following azimuthally varying vectors:\\ 
\begin{equation}
\begin{aligned}
    &\mathbf{e}_{rad}(\varphi) = 
    \mathbf{e}_{H}\cos{\varphi} + \mathbf{e}_{V}\sin{\varphi}~, \\ &\mathbf{e}^{(l)}_{\circlearrowright}(\varphi) =
    (
    \mathbf{e}_{H} + e^{i\frac{\pi}{2}}\mathbf{e}_{V}
    )
    e^{il\varphi}
    ~.
\end{aligned}
\end{equation}

The coincidence probability and the visibility functions calculated in this case depend upon both
$\varphi_{C}$ and $\varphi_{D}$.
\begin{equation}
\begin{aligned}
&C_{(Out)}(\varphi_C,\varphi_D) = {A(\mathbf{r}_C, \mathbf{r}_D)}~, \\ 
&C_{(In)}(\varphi_C,\varphi_D) = \frac{A(\mathbf{r}_C, \mathbf{r}_D)}{2}\left[2 - \cos{2(\varphi_C-\varphi_D)} \right]~, \\ 
&\mathcal{V}(\varphi_C,\varphi_D) = \frac{1}{2}\cos{2(\varphi_C-\varphi_D)}~.
\end{aligned}
\end{equation}
\end{flushleft}

\subsection{Polarisation Projection} 
When polarisation projective measurements are performed, the coincidence probability formula can be derived from equations (\ref{eq: C_in}) and (\ref{eq: C_out}) by eliminating the sum over the independent polarization directions and considering only the directions along which the projection has been performed.
Additionally, we consider that, in this case, $\mathbf{u}_{\alpha}$ and $\mathbf{u}_{\beta}$ are unit-vectors directed along the projective measurement directions.
\begin{align}\label{eq: C_in_polarized}
    C_{(In)}(\mathbf{r}_{C}; \mathbf{r}_{D}) = &
    \frac{A(\mathbf{r}_C, \mathbf{r}_D)}{4} \left| [\mathbf{u}_\alpha \cdot \mathbf{e}_{\sigma_A}(r_C,\varphi_C)] [\mathbf{u}_\beta \cdot \mathbf{e}_{\sigma_B}(r_D,\varphi_D)] - [\mathbf{u}_\alpha \cdot \mathbf{e}_{\sigma_B}(r_C,\varphi_C)] [\mathbf{u}_\beta \cdot \mathbf{e}_{\sigma_A}(r_D,\varphi_D)] \right|^2~,
\end{align}
\begin{align}\label{eq: C_out_polarized}
    C_{(Out)}(\mathbf{r}_{C}; \mathbf{r}_{D}) = &
    \frac{A(\mathbf{r}_C, \mathbf{r}_D)}{4}   \left\{ \left| [\mathbf{u}_\alpha \cdot \mathbf{e}_{\sigma_A}(r_C,\varphi_C)] [\mathbf{u}_\beta \cdot \mathbf{e}_{\sigma_B}(r_D,\varphi_D)] \right|^2 + \left| [\mathbf{u}_\alpha \cdot \mathbf{e}_{\sigma_B}(r_C,\varphi_C)] [\mathbf{u}_\beta \cdot \mathbf{e}_{\sigma_A}(r_D,\varphi_D)] \right|^2 \right\}~.
\end{align}

The computed coincidence probability and visibility distributions are reported for 3 different H-H, H-V, and H-A polariser settings. We remark that $\mathbf{e}_{A} = (\mathbf{e}_{H} - \mathbf{e}_{V})/\sqrt{2}$ is the unit-vector associated with anti-diagonal ($A$) direction.
\begin{equation}\label{C_in_out_HH}
\begin{aligned}
C^{H,H}_{(In)}(\mathbf{r}_C; \mathbf{r}_D) &= \frac{A(\mathbf{r}_C, \mathbf{r}_D)}{4}  \left| H_{\sigma_A}(r_C,\varphi_C) H_{\sigma_B}(r_D,\varphi_D) - H_{\sigma_B}(r_C,\varphi_C) H_{\sigma_A}(r_D,\varphi_D) \right|^2~, \\
C^{H,H}_{(Out)}(\mathbf{r}_C; \mathbf{r}_D) &= \frac{A(\mathbf{r}_C, \mathbf{r}_D)}{4}  \left\{ \left| H_{\sigma_A}(r_C,\varphi_C) H_{\sigma_B}(r_D,\varphi_D) \right|^2 + \left| H_{\sigma_B}(r_C,\varphi_C) H_{\sigma_A}(r_D,\varphi_D) \right|^2 \right\}~,
\end{aligned}
\end{equation}

\begin{equation}\label{C_inout_HV}
\begin{aligned}
C^{H,V}_{(In)}(\mathbf{r}_C; \mathbf{r}_D) =& \frac{A(\mathbf{r}_C, \mathbf{r}_D)}{4} \left| H_{\sigma_A}(r_C,\varphi_C) V_{\sigma_B}(r_D,\varphi_D) - H_{\sigma_B}(r_C,\varphi_C) V_{\sigma_A}(r_D,\varphi_D) \right|^2~, \\
C^{H,V}_{(Out)}(\mathbf{r}_C; \mathbf{r}_D) =& \frac{A(\mathbf{r}_C, \mathbf{r}_D)}{4}
\left\{ \left| 
H_{\sigma_A}(r_C,\varphi_C) 
V_{\sigma_B}(r_D,\varphi_D) \right|^2 + \left| H_{\sigma_B}(r_C,\varphi_C) 
V_{\sigma_A}(r_D,\varphi_D) \right|^2 \right\}~,
\end{aligned}
\end{equation}

\begin{equation}\label{C_inout_HA}
\begin{aligned}
C^{H,A}_{(In)}(\mathbf{r}_C; \mathbf{r}_D) =& \frac{A(\mathbf{r}_C, \mathbf{r}_D)}{8} \left| 
H_{\sigma_A}(r_C,\varphi_C) 
[H_{\sigma_B}(r_D,\varphi_D)-
V_{\sigma_B}(r_D,\varphi_D)] - 
H_{\sigma_B}(r_C,\varphi_C) 
[H_{\sigma_A}(r_D,\varphi_D)-
V_{\sigma_A}(r_D,\varphi_D)] \right|^2~, \\ 
C^{H,A}_{(Out)}(\mathbf{r}_C; \mathbf{r}_D) =& \frac{A(\mathbf{r}_C, \mathbf{r}_D)}{8}
\left\{ \left| 
H_{\sigma_A}(r_C,\varphi_C)
[H_{\sigma_B}(r_D,\varphi_D)-
V_{\sigma_B}(r_D,\varphi_D)] \right|^2 + \left| H_{\sigma_B}(r_C,\varphi_C) 
[H_{\sigma_A}(r_D,\varphi_D)-
V_{\sigma_A}(r_D,\varphi_D)] \right|^2 \right\}~.
\end{aligned}
\end{equation}

Calculation for three cases of interest are performed in the following sections. The calculated coincidence probability and visibility distributions are presented for three different polariser settings.\\

\paragraph{Case: $vv1=rad$ and $vv2=\pi$}
\begin{flushleft}
In H-H and H-V polarisation settings, the visibility map does not depend on the azimuthal angles $\varphi_{C}$ and $\varphi_{D}$, 
and depends only on the $\varphi_{D}$ angle for the H-A configuration.
\begin{equation}
\left.
\begin{aligned}
&C_{(Out)}^{H,H}(\varphi_{C}; \varphi_{D}) =\frac{A(\mathbf{r}_C, \mathbf{r}_D)}{2}\cos^{2}{\varphi_{C}}\cos^{2}{\varphi_{D}}\\
&C_{(In)}^{H,H}(\varphi_{C}; \varphi_{D}) = 0
\end{aligned}
\right\rbrace \quad
\mathcal{V}_{HH}(\varphi_{C}; \varphi_{D}) = 1~,
\end{equation}
\begin{equation}
\left.
\begin{aligned}
&C_{(Out)}^{H,V}(\varphi_{C}; \varphi_{D})=\frac{A(\mathbf{r}_C, \mathbf{r}_D)}{2}\cos^{2}{\varphi_{C}}\sin^{2}{\varphi_{D}}^{2}\\
&C_{(In)}^{H,V}(\varphi_{C}; \varphi_{D})=A(\mathbf{r}_C, \mathbf{r}_D)\cos^{2}{\varphi_{C}}\sin^{2}{\varphi_{D}}
\end{aligned}
\right\rbrace \quad
\mathcal{V}_{HV}(\varphi_{C}; \varphi_{D}) = -1~,
\end{equation}
\begin{equation}
\left.
\begin{aligned}
&C_{(Out)}^{H,A}(\varphi_{C}; \varphi_{D})=\frac{A(\mathbf{r}_C, \mathbf{r}_D)}{4}\cos^{2}{\varphi_{C}}\\
&C_{(In)}^{H,A}(\varphi_{C}; \varphi_{D})=\frac{A(\mathbf{r}_C, \mathbf{r}_D)}{2}\cos^{2}{\varphi_{C}}\sin^{2}{\varphi_{D}}
\end{aligned}
\right\rbrace \quad
\mathcal{V}_{HA}(\varphi_{C}; \varphi_{D}) = \cos{2 \varphi_{D}}~.
\end{equation}

\end{flushleft}

\paragraph{Case: $vv1=rad$ and $vv2=(OAM1)_\circlearrowright$}
\begin{flushleft}
As evident from the equations, in this case, the visibility map depends on the azimuthal angles $\varphi_C$ and $\varphi_D$ for the three different polariser settings.
\begin{equation}
\begin{aligned}
&C_{(Out)}^{H,H}(\varphi_{C}; \varphi_{D})=\frac{A(\mathbf{r}_C, \mathbf{r}_D)}{4}\left[
\cos^{2}{\varphi_{C}} + \cos^{2}{\varphi_{D}}\right]~, \\ 
&C_{(In)}^{H,H}(\varphi_{C}; \varphi_{D})=\frac{A(\mathbf{r}_C, \mathbf{r}_D)}{4}\sin^2{(\varphi_C-\varphi_D)}~,\\
&\mathcal{V}_{HH}(\varphi_{C}; \varphi_{D}) = \frac{2\cos{\varphi_{C}}\cos{\varphi_{D}}\cos(\varphi_{C}-\varphi_{D})}{\cos^{2}{\varphi_{C}} + \cos^{2}{\varphi_{D}}}~,
\end{aligned}
\end{equation}

\begin{equation}
\begin{aligned}
&C_{(Out)}^{H,V}(\varphi_{C}; \varphi_{D}) =\frac{A(\mathbf{r}_C, \mathbf{r}_D)}{4}
\left[\cos^{2}{\varphi_{C}} + \sin^{2}{\varphi_{D}}\right]~,\\ 
&C_{(In)}^{H,V}(\varphi_{C}; \varphi_{D})=\frac{A(\mathbf{r}_C, \mathbf{r}_D)}{4}
\left[C_{(Out)}^{H,V}(\varphi_{C}; \varphi_{D})-2\cos{\varphi_{C}}\sin{\varphi_{D}}\sin(\varphi_{C}-\varphi_{D})\right]~, \\ 
&\mathcal{V}_{HV}(\varphi_{C}; \varphi_{D}) = \frac{2\cos(\varphi_{C})\sin(\varphi_{D})\sin(\varphi_{C}-\varphi_{D})}{
\cos^{2}{\varphi_{C}} + \sin^{2}{\varphi_{D}}}~,
\end{aligned}
\end{equation}

\begin{equation}
\begin{aligned}
&C_{(Out)}^{H,A}(\varphi_{C}; \varphi_{D}) =\frac{A(\mathbf{r}_C, \mathbf{r}_D)}{4}\left[2+\cos{2\varphi_{C}} -
\sin{2\varphi_{D}}\right]~,\\ 
&C_{(In)}^{H,A}(\varphi_{C}; \varphi_{D})=\frac{A(\mathbf{r}_C, \mathbf{r}_D)}{4}\left[C_{(Out)}^{H,A}(\varphi_{C}; \varphi_{D})-[\cos{2\varphi_{D}} - \sin{2\varphi_{C}}+\cos{2(\varphi_{C}-\varphi_{D})}]\right]~,\\ 
&\mathcal{V}_{HA}(\varphi_{C}; \varphi_{D}) = \frac{\cos{2\varphi_{D}} - \sin{2\varphi_{C}}+\cos{2(\varphi_{C}-\varphi_{D})}}{2+\cos{2\varphi_{C}} -\sin{2\varphi_{D}}}~.
\end{aligned}
\end{equation}

\end{flushleft}

\section{Experimental details}
Two photons with  wavelength $\lambda = 810$ nm were generated in the state $\ket{\psi} = \ket{H}\ket{V}$ through type-II degenerate collinear Spontaneous Parametric Down-Conversion (SPDC) by pumping a 30 mm-long periodically poled KTP (ppKTP) crystal with a continuous-wave laser at $\lambda = 405$ nm. 
The crystal was temperature phase-matched at 38$^\circ$C to ensure optimal down-conversion efficiency.

The orthogonally polarised photon pairs were separated by a polarising beam splitter (PBS) and subsequently coupled into single-mode fibres (SMFs) to ensure spatial mode filtering into the fundamental Gaussian mode (TEM$_{00}$). 
Upon exiting the fibres, each photon’s polarisation state was tailored using the sequence of a half-wave plate (HWP) followed by a quarter-wave plate (QWP). The photons were then directed onto electronically tunable liquid crystal \textit{q}-plates with topological charge $q=0.5$, which imparted spatial structuring to the photons.

These structured photons were then injected into the two input ports of a symmetric 50:50 beam splitter (BS) to enable quantum interference. 
Temporal distinguishability between the photons was finely controlled by a motorised translation stage, allowing precise adjustment of the optical path delay.

The outputs from each BS port were directed onto distinct regions of a TimePix3 single-photon sensitive camera, enabling space-resolved detection of coincidence events. 
The camera, equipped with a PhotonisIR intensifier (see Refs. \cite{fisher2016timepixcam, TPX3CamSpatialCaracterization, TPX3CamCaracterization} for more information), operates with an overall detection efficiency of approximately $19\%$ at $\lambda = 810$ nm.
The data consisted of time-stamped signals detected at each pixel. Due to the spatial spread of the signal induced by the amplification stage of the intensifier, a clustering algorithm was employed to reconstruct the spatial position of each individual photon hit.
The 2D visibility maps were obtained by dividing the measured images into 28 circular sectors for each mode and computing the coincidence counts between all pairs of sectors.
Where required, polarisation projections were performed by inserting linear polarisers in each output arm prior to detection, allowing further control and manipulation of the spatial correlation patterns.

\begin{figure}
    \centering
    \includegraphics[width=1\linewidth]{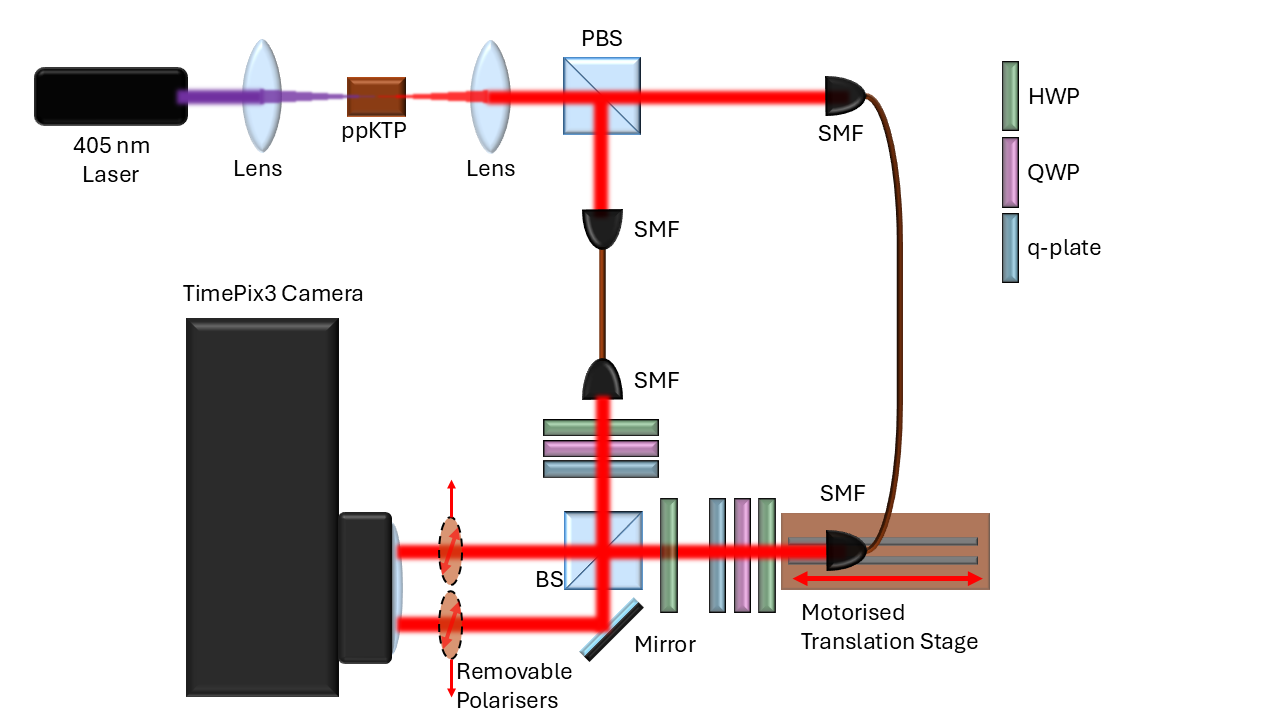}
    \caption{\textbf{Detailed experimental Setup.}
    A 30mm-long ppKTP crystal is pumped with a $\lambda = 405$ nm continuous-wave laser to generate orthogonally polarized photon pairs. 
    These are path separated by a PBS and coupled into single-mode fibers (SMF) for spatial-mode filtering. 
    The input polarization is set using HWP-QWP, followed by q-plates for polarization-dependent spatial structuring.
    The photons then interfere at a 50:50 BS and the outputs are directed to a TimePix3 camera for spatially-resolved measurements.
    }
    \label{setup}
\end{figure}


\begin{thebibliography}{10}
\expandafter\ifx\csname url\endcsname\relax
  \def\url#1{\texttt{#1}}\fi
\expandafter\ifx\csname urlprefix\endcsname\relax\def\urlprefix{URL }\fi
\providecommand{\bibinfo}[2]{#2}
\providecommand{\eprint}[2][]{\url{#2}}

\bibitem{lubin2022photon}
\bibinfo{author}{Lubin, G.}, \bibinfo{author}{Oron, D.},
  \bibinfo{author}{Rossman, U.}, \bibinfo{author}{Tenne, R.} \&
  \bibinfo{author}{Yallapragada, V.~J.}
\newblock \bibinfo{title}{Photon correlations in spectroscopy and microscopy}.
\newblock \emph{\bibinfo{journal}{ACS photonics}} \textbf{\bibinfo{volume}{9}},
  \bibinfo{pages}{2891--2904} (\bibinfo{year}{2022}).

\bibitem{Defi24}
\bibinfo{author}{Defienne, H.} \emph{et~al.}
\newblock \bibinfo{title}{Advances in quantum imaging}.
\newblock \emph{\bibinfo{journal}{Nature Photonics}}
  \textbf{\bibinfo{volume}{18}}, \bibinfo{pages}{1024--1036}
  (\bibinfo{year}{2024}).

\bibitem{bigbelltest}
\bibinfo{author}{{BIG Bell Test Collaboration}}.
\newblock \bibinfo{title}{Challenging local realism with human choices}.
\newblock \emph{\bibinfo{journal}{Nature}} \textbf{\bibinfo{volume}{557}}
  (\bibinfo{year}{2018}).

\bibitem{gilaberte2023experimental}
\bibinfo{author}{Gilaberte~Basset, M.} \emph{et~al.}
\newblock \bibinfo{title}{Experimental analysis of image resolution of quantum
  imaging with undetected light through position correlations}.
\newblock \emph{\bibinfo{journal}{Physical Review A}}
  \textbf{\bibinfo{volume}{108}}, \bibinfo{pages}{052610}
  (\bibinfo{year}{2023}).

\bibitem{genovese2016real}
\bibinfo{author}{Genovese, M.}
\newblock \bibinfo{title}{Real applications of quantum imaging}.
\newblock \emph{\bibinfo{journal}{Journal of Optics}}
  \textbf{\bibinfo{volume}{18}}, \bibinfo{pages}{073002}
  (\bibinfo{year}{2016}).

\bibitem{england2019quantum}
\bibinfo{author}{England, D.~G.}, \bibinfo{author}{Balaji, B.} \&
  \bibinfo{author}{Sussman, B.~J.}
\newblock \bibinfo{title}{Quantum-enhanced standoff detection using correlated
  photon pairs}.
\newblock \emph{\bibinfo{journal}{Physical Review A}}
  \textbf{\bibinfo{volume}{99}}, \bibinfo{pages}{023828}
  (\bibinfo{year}{2019}).

\bibitem{jackson2002optical}
\bibinfo{author}{Jackson, D.}, \bibinfo{author}{Hockney, G.} \&
  \bibinfo{author}{Dowling, J.}
\newblock \bibinfo{title}{Optical communication noise rejection using
  correlated photons}.
\newblock \emph{\bibinfo{journal}{journal of modern optics}}
  \textbf{\bibinfo{volume}{49}}, \bibinfo{pages}{2383--2388}
  (\bibinfo{year}{2002}).

\bibitem{Branczyk11}
\bibinfo{author}{Bra\'{n}czyk, A.~M.}, \bibinfo{author}{Fedrizzi, A.},
  \bibinfo{author}{Stace, T.~M.}, \bibinfo{author}{Ralph, T.~C.} \&
  \bibinfo{author}{White, A.~G.}
\newblock \bibinfo{title}{Engineered optical nonlinearity for quantum light
  sources}.
\newblock \emph{\bibinfo{journal}{Opt. Express}} \textbf{\bibinfo{volume}{19}},
  \bibinfo{pages}{55--65} (\bibinfo{year}{2011}).

\bibitem{Dosseva2016}
\bibinfo{author}{Dosseva, A.}, \bibinfo{author}{Cincio, L.} \&
  \bibinfo{author}{Bra\ifmmode~\acute{n}\else \'{n}\fi{}czyk, A.~M.}
\newblock \bibinfo{title}{Shaping the joint spectrum of down-converted photons
  through optimized custom poling}.
\newblock \emph{\bibinfo{journal}{Phys. Rev. A}} \textbf{\bibinfo{volume}{93}},
  \bibinfo{pages}{013801} (\bibinfo{year}{2016}).

\bibitem{Yesh23}
\bibinfo{author}{Yesharim, O.}, \bibinfo{author}{Pearl, S.},
  \bibinfo{author}{Foley-Comer, J.}, \bibinfo{author}{Juwiler, I.} \&
  \bibinfo{author}{Arie, A.}
\newblock \bibinfo{title}{Direct generation of spatially entangled qudits using
  quantum nonlinear optical holography}.
\newblock \emph{\bibinfo{journal}{Science Advances}}
  \textbf{\bibinfo{volume}{9}}, \bibinfo{pages}{eade7968}
  (\bibinfo{year}{2023}).

\bibitem{Valencia2007}
\bibinfo{author}{Valencia, A.}, \bibinfo{author}{Cer\'e, A.},
  \bibinfo{author}{Shi, X.}, \bibinfo{author}{Molina-Terriza, G.} \&
  \bibinfo{author}{Torres, J.~P.}
\newblock \bibinfo{title}{Shaping the waveform of entangled photons}.
\newblock \emph{\bibinfo{journal}{Phys. Rev. Lett.}}
  \textbf{\bibinfo{volume}{99}}, \bibinfo{pages}{243601}
  (\bibinfo{year}{2007}).

\bibitem{Mair2021}
\bibinfo{author}{Mair, A.}, \bibinfo{author}{Vaziri, A.} \&
  \bibinfo{author}{Weihs, G. e.~a.}
\newblock \bibinfo{title}{Entanglement of the orbital angular momentum states
  of photons}.
\newblock \emph{\bibinfo{journal}{Nature}} \textbf{\bibinfo{volume}{412}},
  \bibinfo{pages}{313–316} (\bibinfo{year}{2001}).

\bibitem{Kovlakov2017}
\bibinfo{author}{Kovlakov, E.~V.}, \bibinfo{author}{Bobrov, I.~B.},
  \bibinfo{author}{Straupe, S.~S.} \& \bibinfo{author}{Kulik, S.~P.}
\newblock \bibinfo{title}{Spatial bell-state generation without transverse mode
  subspace postselection}.
\newblock \emph{\bibinfo{journal}{Phys. Rev. Lett.}}
  \textbf{\bibinfo{volume}{118}}, \bibinfo{pages}{030503}
  (\bibinfo{year}{2017}).

\bibitem{Kovlakov2018}
\bibinfo{author}{Kovlakov, E.~V.}, \bibinfo{author}{Straupe, S.~S.} \&
  \bibinfo{author}{Kulik, S.~P.}
\newblock \bibinfo{title}{Quantum state engineering with twisted photons via
  adaptive shaping of the pump beam}.
\newblock \emph{\bibinfo{journal}{Phys. Rev. A}} \textbf{\bibinfo{volume}{98}},
  \bibinfo{pages}{060301} (\bibinfo{year}{2018}).

\bibitem{Boucher2021}
\bibinfo{author}{Boucher, P.}, \bibinfo{author}{Defienne, H.} \&
  \bibinfo{author}{Gigan, S.}
\newblock \bibinfo{title}{Engineering spatial correlations of entangled photon
  pairs by pump beam shaping}.
\newblock \emph{\bibinfo{journal}{Opt. Lett.}} \textbf{\bibinfo{volume}{46}},
  \bibinfo{pages}{4200--4203} (\bibinfo{year}{2021}).

\bibitem{Unternahrer:18}
\bibinfo{author}{Untern\"{a}hrer, M.}, \bibinfo{author}{Bessire, B.},
  \bibinfo{author}{Gasparini, L.}, \bibinfo{author}{Perenzoni, M.} \&
  \bibinfo{author}{Stefanov, A.}
\newblock \bibinfo{title}{Super-resolution quantum imaging at the heisenberg
  limit}.
\newblock \emph{\bibinfo{journal}{Optica}} \textbf{\bibinfo{volume}{5}},
  \bibinfo{pages}{1150--1154} (\bibinfo{year}{2018}).

\bibitem{Fran21}
\bibinfo{author}{Francesconi, S.} \emph{et~al.}
\newblock \bibinfo{title}{Anyonic two-photon statistics with a semiconductor
  chip}.
\newblock \emph{\bibinfo{journal}{ACS Photonics}} \textbf{\bibinfo{volume}{8}},
  \bibinfo{pages}{2764--2769} (\bibinfo{year}{2021}).

\bibitem{Defienne2024}
\bibinfo{author}{Verni\`ere, C.} \& \bibinfo{author}{Defienne, H.}
\newblock \bibinfo{title}{Hiding images in quantum correlations}.
\newblock \emph{\bibinfo{journal}{Phys. Rev. Lett.}}
  \textbf{\bibinfo{volume}{133}}, \bibinfo{pages}{093601}
  (\bibinfo{year}{2024}).

\bibitem{DAmb16}
\bibinfo{author}{D'Ambrosio, V.} \emph{et~al.}
\newblock \bibinfo{title}{Entangled vector vortex beams}.
\newblock \emph{\bibinfo{journal}{Physical ReviewA}}
  \textbf{\bibinfo{volume}{94}}, \bibinfo{pages}{030304}
  (\bibinfo{year}{2016}).

\bibitem{Gao_2024}
\bibinfo{author}{Gao, X.} \emph{et~al.}
\newblock \bibinfo{title}{Full spatial characterization of entangled structured
  photons}.
\newblock \emph{\bibinfo{journal}{Phys. Rev. Lett.}}
  \textbf{\bibinfo{volume}{132}}, \bibinfo{pages}{063802}
  (\bibinfo{year}{2024}).

\bibitem{Cameron2024}
\bibinfo{author}{Cameron, P.}, \bibinfo{author}{Courmea, B.},
  \bibinfo{author}{Faccio, D.} \& \bibinfo{author}{Defienne, H.}
\newblock \bibinfo{title}{Shaping the spatial correlations of entangled photon
  pairs}.
\newblock \emph{\bibinfo{journal}{J. Phys. Photonics}}
  \textbf{\bibinfo{volume}{6}}, \bibinfo{pages}{033001} (\bibinfo{year}{2024}).

\bibitem{3xSPDC}
\bibinfo{author}{Chang, C. W.~S.} \emph{et~al.}
\newblock \bibinfo{title}{Observation of three-photon spontaneous parametric
  down-conversion in a superconducting parametric cavity}.
\newblock \emph{\bibinfo{journal}{Phys. Rev. X}} \textbf{\bibinfo{volume}{10}},
  \bibinfo{pages}{011011} (\bibinfo{year}{2020}).

\bibitem{bouchard2020two}
\bibinfo{author}{Bouchard, F.} \emph{et~al.}
\newblock \bibinfo{title}{Two-photon interference: the {Hong--Ou--Mandel}
  effect}.
\newblock \emph{\bibinfo{journal}{Reports on Progress in Physics}}
  \textbf{\bibinfo{volume}{84}}, \bibinfo{pages}{012402}
  (\bibinfo{year}{2020}).

\bibitem{kwiat1992observation}
\bibinfo{author}{Kwiat, P.~G.}, \bibinfo{author}{Steinberg, A.~M.} \&
  \bibinfo{author}{Chiao, R.~Y.}
\newblock \bibinfo{title}{Observation of a ‘‘quantum eraser’’: A
  revival of coherence in a two-photon interference experiment}.
\newblock \emph{\bibinfo{journal}{Physical Review A}}
  \textbf{\bibinfo{volume}{45}}, \bibinfo{pages}{7729} (\bibinfo{year}{1992}).

\bibitem{walborn2003multimode}
\bibinfo{author}{Walborn, S.}, \bibinfo{author}{De~Oliveira, A.},
  \bibinfo{author}{P{\'a}dua, S.} \& \bibinfo{author}{Monken, C.}
\newblock \bibinfo{title}{Multimode {Hong-Ou-Mandel} interference}.
\newblock \emph{\bibinfo{journal}{Physical review letters}}
  \textbf{\bibinfo{volume}{90}}, \bibinfo{pages}{143601}
  (\bibinfo{year}{2003}).

\bibitem{nagali2009optimal}
\bibinfo{author}{Nagali, E.} \emph{et~al.}
\newblock \bibinfo{title}{Optimal quantum cloning of orbital angular momentum
  photon qubits through {Hong--Ou--Mandel} coalescence}.
\newblock \emph{\bibinfo{journal}{Nature Photonics}}
  \textbf{\bibinfo{volume}{3}}, \bibinfo{pages}{720--723}
  (\bibinfo{year}{2009}).

\bibitem{di2010measurement}
\bibinfo{author}{Di~Lorenzo~Pires, H.}, \bibinfo{author}{Florijn, H.} \&
  \bibinfo{author}{Van~Exter, M.}
\newblock \bibinfo{title}{Measurement of the spiral spectrum of entangled
  two-photon states}.
\newblock \emph{\bibinfo{journal}{Physical review letters}}
  \textbf{\bibinfo{volume}{104}}, \bibinfo{pages}{020505}
  (\bibinfo{year}{2010}).

\bibitem{karimi2014exploring}
\bibinfo{author}{Karimi, E.} \emph{et~al.}
\newblock \bibinfo{title}{Exploring the quantum nature of the radial degree of
  freedom of a photon via {Hong-Ou-Mandel} interference}.
\newblock \emph{\bibinfo{journal}{Physical Review A}}
  \textbf{\bibinfo{volume}{89}}, \bibinfo{pages}{013829}
  (\bibinfo{year}{2014}).

\bibitem{Schi2024}
\bibinfo{author}{Schiano, C.} \emph{et~al.}
\newblock \bibinfo{title}{Engineering quantum states from a spatially
  structured quantum eraser}.
\newblock \emph{\bibinfo{journal}{Science Advances}}
  \textbf{\bibinfo{volume}{10}}, \bibinfo{pages}{eadm9278}
  (\bibinfo{year}{2024}).

\bibitem{Gao25}
\bibinfo{author}{Gao, X.}, \bibinfo{author}{Paneru, D.},
  \bibinfo{author}{Di~Colandrea, F.}, \bibinfo{author}{Zhang, Y.} \&
  \bibinfo{author}{Karimi, E.}
\newblock \bibinfo{title}{Generation of the complete bell basis via
  {Hong-Ou-Mandel} interference of vector modes}.
\newblock \emph{\bibinfo{journal}{Phys. Rev. A}}
  \textbf{\bibinfo{volume}{112}}, \bibinfo{pages}{012215}
  (\bibinfo{year}{2025}).

\bibitem{bornman2019ghost}
\bibinfo{author}{Bornman, N.} \emph{et~al.}
\newblock \bibinfo{title}{Ghost imaging using entanglement-swapped photons}.
\newblock \emph{\bibinfo{journal}{npj Quantum Information}}
  \textbf{\bibinfo{volume}{5}}, \bibinfo{pages}{63} (\bibinfo{year}{2019}).

\bibitem{zhang2017simultaneous}
\bibinfo{author}{Zhang, Y.} \emph{et~al.}
\newblock \bibinfo{title}{Simultaneous entanglement swapping of multiple
  orbital angular momentum states of light}.
\newblock \emph{\bibinfo{journal}{Nature communications}}
  \textbf{\bibinfo{volume}{8}}, \bibinfo{pages}{632} (\bibinfo{year}{2017}).

\bibitem{zhang2016engineering}
\bibinfo{author}{Zhang, Y.} \emph{et~al.}
\newblock \bibinfo{title}{Engineering two-photon high-dimensional states
  through quantum interference}.
\newblock \emph{\bibinfo{journal}{Science advances}}
  \textbf{\bibinfo{volume}{2}}, \bibinfo{pages}{e1501165}
  (\bibinfo{year}{2016}).

\bibitem{Casalengua_2024_VortexCorrelations}
\bibinfo{author}{Zubizarreta~Casalengua, E.} \& \bibinfo{author}{Laussy, F.~P.}
\newblock \bibinfo{title}{Spatial correlations of vortex quantum states}.
\newblock \emph{\bibinfo{journal}{arXiv preprint arXiv:2402.01627}}
  (\bibinfo{year}{2024}).

\bibitem{Photons_Atoms}
\bibinfo{author}{C.~Cohen-Tannoudji, G.~G., J. Dupont-Roc}.
\newblock \emph{\bibinfo{title}{Photons and Atoms: Basic Processes and
  Applications}} (\bibinfo{publisher}{John Wiley \& Sons},
  \bibinfo{address}{New York}, \bibinfo{year}{1992}).

\bibitem{weihs1996two}
\bibinfo{author}{Weihs, G.}, \bibinfo{author}{Reck, M.},
  \bibinfo{author}{Weinfurter, H.} \& \bibinfo{author}{Zeilinger, A.}
\newblock \bibinfo{title}{Two-photon interference in optical fiber multiports}.
\newblock \emph{\bibinfo{journal}{Physical Review A}}
  \textbf{\bibinfo{volume}{54}}, \bibinfo{pages}{893} (\bibinfo{year}{1996}).

\bibitem{Gard:13}
\bibinfo{author}{Gard, B.~T.}, \bibinfo{author}{Cross, R.~M.},
  \bibinfo{author}{Anisimov, P.~M.}, \bibinfo{author}{Lee, H.} \&
  \bibinfo{author}{Dowling, J.~P.}
\newblock \bibinfo{title}{Quantum random walks with multiphoton interference
  and high-order correlation functions}.
\newblock \emph{\bibinfo{journal}{J. Opt. Soc. Am. B}}
  \textbf{\bibinfo{volume}{30}}, \bibinfo{pages}{1538--1545}
  (\bibinfo{year}{2013}).

\bibitem{12-Photon-Entanglement}
\bibinfo{author}{Zhong, H.-S.} \emph{et~al.}
\newblock \bibinfo{title}{12-photon entanglement and scalable scattershot boson
  sampling with optimal entangled-photon pairs from parametric
  down-conversion}.
\newblock \emph{\bibinfo{journal}{Phys. Rev. Lett.}}
  \textbf{\bibinfo{volume}{121}}, \bibinfo{pages}{250505}
  (\bibinfo{year}{2018}).

\bibitem{20xBoson_Sampling}
\bibinfo{author}{Wang, H.} \emph{et~al.}
\newblock \bibinfo{title}{Boson sampling with 20 input photons and a 60-mode
  interferometer in a $1{0}^{14}$-dimensional {Hilbert} space}.
\newblock \emph{\bibinfo{journal}{Phys. Rev. Lett.}}
  \textbf{\bibinfo{volume}{123}}, \bibinfo{pages}{250503}
  (\bibinfo{year}{2019}).

\bibitem{jaouni2025tutorial}
\bibinfo{author}{Jaouni, T.}, \bibinfo{author}{Gu, X.}, \bibinfo{author}{Krenn,
  M.}, \bibinfo{author}{D'Errico, A.} \& \bibinfo{author}{Karimi, E.}
\newblock \bibinfo{title}{Tutorial: {Hong-Ou-Mandel} interference with
  structured photons}.
\newblock \emph{\bibinfo{journal}{arXiv preprint arXiv:2501.14961}}
  (\bibinfo{year}{2025}).

\bibitem{Intro_QED}
\bibinfo{author}{C.~Cohen-Tannoudji, G.~G., J. Dupont-Roc}.
\newblock \emph{\bibinfo{title}{Photons and Atoms: Introduction to Quantum
  Electrodynamics}} (\bibinfo{publisher}{John Wiley \& Sons},
  \bibinfo{address}{New York}, \bibinfo{year}{1997}).

\bibitem{fisher2016timepixcam}
\bibinfo{author}{Fisher-Levine, M.} \& \bibinfo{author}{Nomerotski, A.}
\newblock \bibinfo{title}{Timepixcam: a fast optical imager with
  time-stamping}.
\newblock \emph{\bibinfo{journal}{Journal of Instrumentation}}
  \textbf{\bibinfo{volume}{11}}, \bibinfo{pages}{C03016}
  (\bibinfo{year}{2016}).

\bibitem{TPX3CamSpatialCaracterization}
\bibinfo{author}{Ianzano, C.} \emph{et~al.}
\newblock \bibinfo{title}{Fast camera spatial characterization of photonic
  polarization etanglement}.
\newblock \emph{\bibinfo{journal}{Scientific Reports}}
  \textbf{\bibinfo{volume}{10}} (\bibinfo{year}{2020}).

\bibitem{TPX3CamCaracterization}
\bibinfo{author}{Vidyapin, V.}, \bibinfo{author}{Zhang, Y.},
  \bibinfo{author}{Englend, D.} \& \bibinfo{author}{Sussman, B.}
\newblock \bibinfo{title}{Characterisation of a single photon event camera for
  quantum imaging}.
\newblock \emph{\bibinfo{journal}{Scientific Reports}}
  \textbf{\bibinfo{volume}{13}} (\bibinfo{year}{2023}).

\end{thebibliography}
\end{document}